\documentclass{article}

\usepackage{arxiv}

\usepackage[utf8]{inputenc} 
\usepackage[T1]{fontenc}    
\usepackage{hyperref}       
\usepackage{url}            
\usepackage{booktabs}       
\usepackage{amsfonts}       
\usepackage{nicefrac}       
\usepackage{microtype}      
\usepackage{lipsum}		
\usepackage{graphicx}

\usepackage{graphicx}
\usepackage{amsmath}
\usepackage{ragged2e}

\begin{document}

	\title{Enhanced TNSA Ion Acceleration via Optical Confinement and Geometric Plasma Focusing in Annular Sector Targets}

\author{
		\href{https://orcid.org/0000-0002-3538-7933}{\includegraphics[scale=0.06]{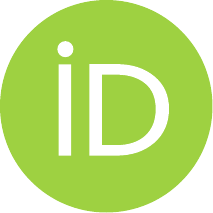}\hspace{1mm}Mohammad Rezaei-Pandari}\\
			Department of physics\\
		Iran University of Science and Technology\\
		Narmak, Tehran, Iran \\
		\texttt{rezaei.m.p@gmail.com} \\
	\AND
		\href{https://orcid.org/0000-0003-2218-072X}{\includegraphics[scale=0.06]{orcid.pdf}\hspace{1mm}Mahdi Shayganmanesh}\\
		Department of physics\\
		Iran University of Science and Technology\\
		Narmak, Tehran, Iran \\
			\AND
			\href{https://orcid.org/0000-0003-0345-2259}{\includegraphics[scale=0.06]{orcid.pdf}\hspace{1mm}Mohammad Hossein Mahdieh}\\
			Department of physics\\
			Iran University of Science and Technology\\
			Narmak, Tehran, Iran \\
		}
	

\maketitle 

\renewcommand{\shorttitle}{Enhanced TNSA Ion Acceleration via Optical Confinement}

\hypersetup{
	pdftitle={Enhanced TNSA Ion Acceleration via Optical Confinement},
	pdfsubject={q-bio.NC, q-bio.QM},
	pdfauthor={Mohammad Rezaei-Pandari, Mahdi Shayganmanesh,Mohammad Hossein Mahdieh},
	pdfkeywords={Laser-driven ion acceleration, Target Normal Sheath Acceleration (TNSA), Annular sector target,Optical confinement,Geometric plasma focusing},
}

\begin{abstract}
	\justifying
	
		Enhancing the conversion efficiency and maximum energy of laser-driven ion beams is a critical challenge for applications in hadron therapy and high-energy density physics. In this work, we present a comprehensive two-dimensional Particle-In-Cell (PIC) simulation study comparing Target Normal Sheath Acceleration (TNSA) from standard flat foils and novel annular sector (C-shaped) targets. Under identical ultra-intense laser irradiation ($a_0=10$, $\tau=25\,\mathrm{fs}$), the annular sector geometry demonstrates a substantial enhancement in acceleration performance driven by two synergistic mechanisms: electromagnetic cavity confinement and geometric plasma focusing. Our analysis reveals that the target void acts as an optical trap, sustaining oscillating electromagnetic fields for over $300\,\mathrm{fs}$ via multiple internal reflections. This confinement results in a total laser energy absorption of $49\%$ (compared to $16\%$ for flat targets), which yields a peak electron temperature of $5.1\,\mathrm{MeV}$—more than double the $2.2\,\mathrm{MeV}$ observed in flat targets. Furthermore, phase space diagnostics confirm that ion bunches accelerated from the converging cavity walls superimpose at the geometric center, creating a localized high-density focal spot. Consequently, the annular target increases the proton cut-off energy to $22\,\mathrm{MeV}$ (vs. $12\,\mathrm{MeV}$ for flat targets) and boosts Carbon ion energies beyond $60\,\mathrm{MeV}$. These findings establish that tailoring target curvature to exploit optical trapping and geometric focusing offers a robust pathway for developing compact, high-efficiency laser-ion sources.
\end{abstract}
	
	\keywords{Laser-driven ion acceleration, Target Normal Sheath Acceleration (TNSA), Annular sector target,Optical confinement,Geometric plasma focusing}

	\section{Introduction}
	The advent of high-power lasers, including access to petawatt-class systems, has opened unprecedented avenues for exploring laser-matter interactions. These advancements enable unique applications such as the generation of high-brightness betatron radiation sources\cite{hojbota2023high}, energetic femtosecond electron bunches\cite{aniculaesei2023acceleration}, and high-energy ion sources\cite{bisesto2020ultrafast,wang2025enhanced}. These capabilities hold immense potential for transformative applications in advanced imaging techniques\cite{zhang2025high-brightness,guo2025preclinical}, terahertz generation\cite{pak2023multi-millijoule,rezaei-pandari2024investigation}, and fundamental physics research\cite{tajima2012fundamental}.

	Laser-driven ion acceleration is a rapidly evolving field with promising applications in areas like cancer therapy\cite{linz12laser-driven,ma2006development}, proton radiography\cite{Simpson_2021,huang2025characterization,borghesi2008laser-driven}, and inertial confinement fusion\cite{badziak2021laser-driven,tikhonchuk2018physics}. This technique leverages ultrafast, high-peak-power lasers to generate extremely strong electric fields within plasma, accelerating ions to relativistic velocities over femtosecond-to-picosecond timescales and sub-millimeter distances. Compared to conventional accelerators, this process offers significant advantages, including reduced complexity, size, and cost\cite{badziak2018laser-driven,cho2025laser-powered}. However, standard TNSA from flat foils typically produces highly divergent ion beams with broad energy spreads, which limits their transport and coupling efficiency for applications requiring localized energy deposition.

	Several mechanisms contribute to laser-driven ion acceleration, such as target normal sheath acceleration (TNSA)\cite{borghesi2019ion,gizzi2017new,roth2013ion,costa2020dependence,bisesto2020simultaneous,iwawaki2021backward} and backward plasma acceleration (BPA)\cite{roshan2009backward}. The efficiency of these mechanisms is highly influenced by factors like target composition, geometry, and laser parameters\cite{hadjikyriacou2023novel,gizzi2017new,schreiber2014optimization,permogorov2021target,loughran2023automated,ferri2019enhanced,Ter-Avetisyan_2023,Badziak_2021}. TNSA is a robust and widely studied mechanism for accelerating protons and ions from solid targets irradiated by high-power lasers. In TNSA, an intense laser pulse rapidly ionizes the target material, generating a population of highly energetic ("hot") electrons. These hot electrons propagate through the target and escape its rear surface, creating a powerful electrostatic field—the sheath field—due to charge separation. This sheath field then efficiently accelerates ions (typically protons from surface contaminants) normal to the target's rear surface\cite{borghesi2019ion,roth2013ion,yin2006gev}.
	
	Ongoing research in TNSA focuses on optimizing target geometry to address these limitations\cite{ferri2019enhanced,gizzi2021enhanced,pazzaglia2020theoretical,tazes2024efficient,vallires2021enhanced,zou2017laser-driven}. Strategies generally fall into two categories: enhancing laser absorption and shaping the ion beam. To boost absorption, structured targets such as gratings, nanowires, and foams have been employed to increase the hot electron yield\cite{Ceccotti2013,vallires2021enhanced,GOLOVIN2020,Roycroft2020}. Conversely, to improve beam quality and collimation, curved targets like hemi-shells and cones have been investigated to spatially focus the accelerating sheath \cite{Morace2022,Kluge2012}. Recently, more complex micro-structures have emerged. For instance, Khan et al. demonstrated a four-fold enhancement using a rectangularly grooved target \cite{khan2023enhanced}, while multi-channel wire targets have shown improved field generation \cite{zou2019enhancement}. These advancements suggest that combining optical confinement structures with focusing geometries could yield superior acceleration performance.

	Despite these advancements, a critical challenge remains: how to further improve the efficiency of converting laser energy into ion acceleration. In this study, we propose and investigate a novel annular sector (C-shaped) target design that integrates these concepts. Unlike simple flat or curved foils, this geometry creates a semi-enclosed cavity. We hypothesize that this design will induce a dual-enhancement mechanism: (1) acting as an optical trap to recirculate the laser pulse to sustain electron heating, and (2) utilizing geometric plasma focusing to concentrate the energetic ions at a focal point. Using 2D PIC simulations, we demonstrate that this synergy leads to significantly higher electron temperatures and ion cut-off energies compared to planar equivalents. This paper is structured as follows: we first detail the simulation setup, followed by a comprehensive presentation and discussion of our results, and conclude with a summary of our findings.

	\section{Simulation Setup}
	
	Two-dimensional particle-in-cell (PIC) simulations were performed using the open-source code \textsc{Smilei}~\cite{derouillat2018smilei}. The simulation domain was defined with dimensions $L_x \times L_y = 50\,\mu\mathrm{m} \times 30\,\mu\mathrm{m}$. The spatial domain was discretized with a resolution of $\Delta x = \Delta y = \lambda_0/32 \approx 25\,\mathrm{nm}$ to strictly resolve the plasma skin depth.
	
	A linearly polarized Gaussian laser pulse with a central wavelength $\lambda_0 = 800\,\mathrm{nm}$, FWHM duration $\tau = 25\,\mathrm{fs}$, and a normalized vector potential $a_0 = 10$ was initialized propagating in the $+x$ direction. The laser was focused to a spot size of $w_0 = 4\,\mu\mathrm{m}$ at the focal plane $x_f = 18\,\mu\mathrm{m}$.
	
	We investigated two distinct target geometries: a flat foil and an annular sector (cavity) target. The bulk target was composed of a plastic (CH) plasma. The ion densities were initialized as $n_C = n_H = 26.44\,n_{cr}$. Corresponding to an initial ionization state of fully ionized hydrogen ($H^+$) and triply ionized carbon ($C^{3+}$), the resulting initial electron density was set to $n_e \approx 106\,n_{cr}$. To account for the finite contrast of the laser pulse and the resulting pre-expansion of the target surface, a pre-plasma profile was initialized at the density interfaces. This pre-plasma follows a hyperbolic tangent gradient with a characteristic decay length $\delta = 0.1\,\mu\mathrm{m}$.
	
	The flat target had dimensions of $1\,\mu\mathrm{m}$ thickness and $27\,\mu\mathrm{m}$ width, while the annular sector target is centered at the laser focal plane ($x_c = 18\,\mu\mathrm{m}$, $y_c = 15\,\mu\mathrm{m}$) and features an inner radius $R_{in} = 6\,\mu\mathrm{m}$ and a thickness of $1\,\mu\mathrm{m}$. The geometry forms a $300^{\circ}$ C-shaped structure with a $60^{\circ}$ angular void facing the incoming laser (defined by angles $150^{\circ} < \theta < 210^{\circ}$ relative to the laser axis). This configuration allows the laser pulse to propagate through the vacuum center before interacting with the pre-plasma layer on the concave back surface.
	
	The simulation was initialized with 25 macro-particles per cell per species. While Carbon ions were initialized with a charge state $Z=3$, the tunnel ionization module was enabled to model further ionization dynamics ($C^{3+} \rightarrow C^{6+}$) in the high-field region.

	\begin{figure}
		\centering
		\includegraphics[width=0.7\linewidth]{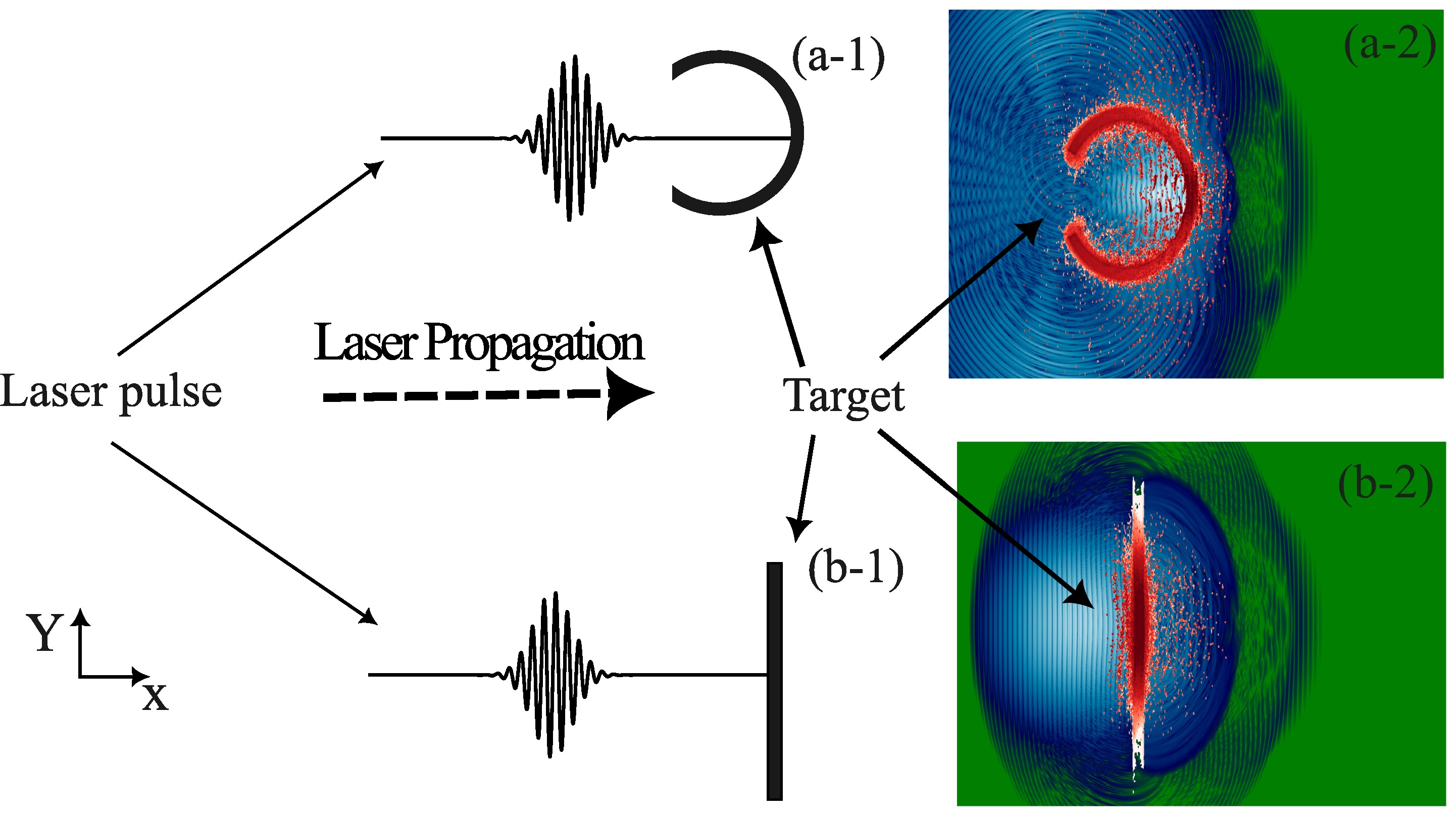}
		\caption{Schematic of laser-solid target interaction.  (a-1) and (b-1) depict a laser pulse propagating in the x-direction and interacting with a annular sector and a flat target, respectively.  (a-2) and (b-2) show simulation snapshots of the electric field intensity and electron distribution during these interactions.}
		\label{fig:fig1}
	\end{figure}
	
	\section{Results: Temporal Evolution of Interaction}
	
	The temporal evolution of the laser-target interaction is presented in Figs.~\ref{fig:fig2} and \ref{fig:fig3}. These figures display 2D spatial maps of ion energy density (top panels) and electron energy density (bottom panels). Overlaid on these density maps are the laser intensity profiles (green-blue scale) and 1D lineouts of the electric fields along the central axis ($y=0$). Specifically, the lineout on the ion maps depicts the transverse laser field ($E_y$), while the lineout on the electron maps depicts the longitudinal electrostatic field ($E_x$).
	
	\subsection{Early-Stage Interaction ($t=100$--$140\,\mathrm{fs}$)}
	
\begin{figure}
		\centering
		\includegraphics[width=0.9\linewidth]{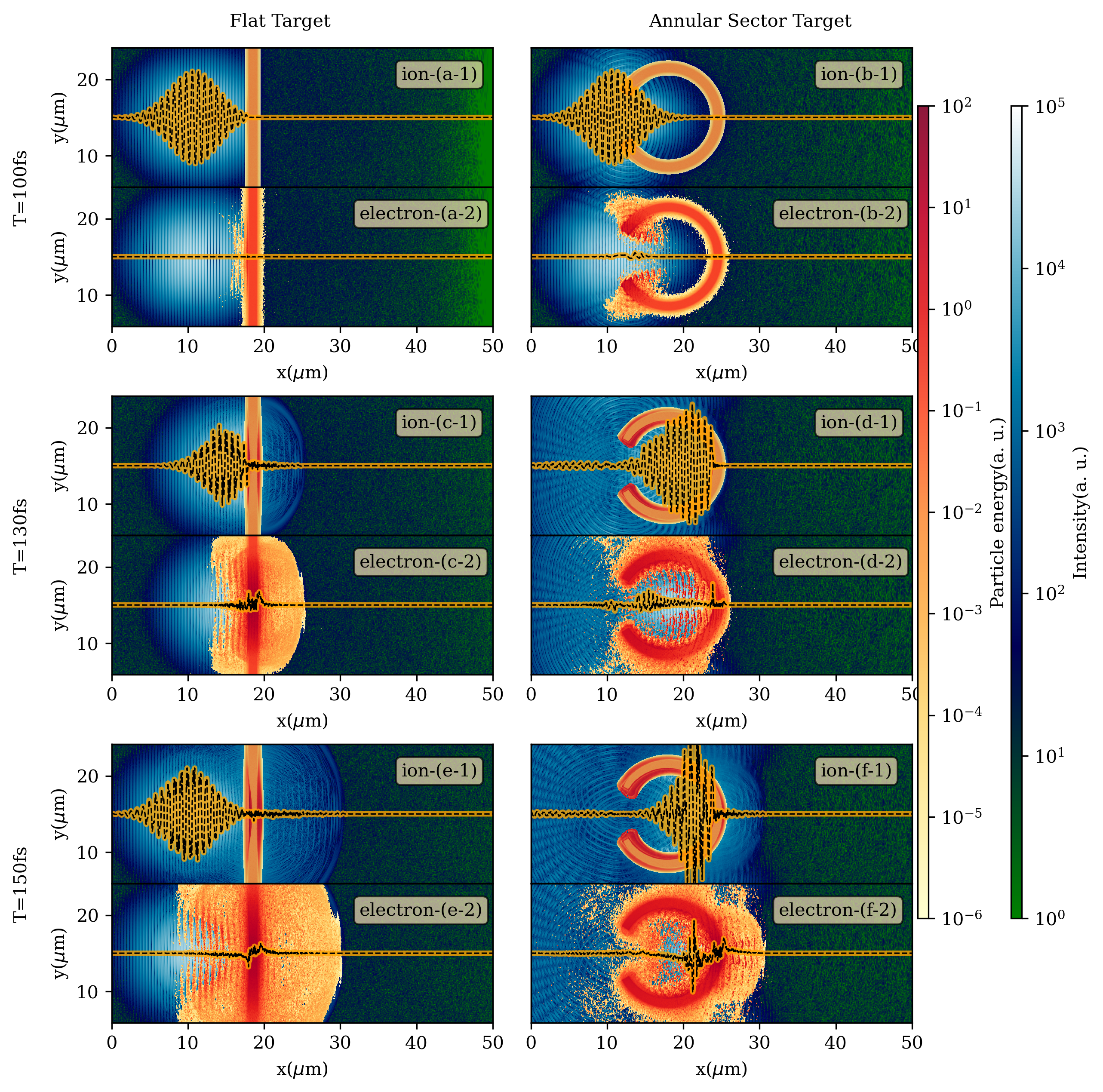}
		\caption{Comparison of ion and electron energy densities in laser-solid target interactions for flat and annular sector targets. The left column displays results for the flat target, while the right column shows the annular sector target. In each subfigure, the top panel (labeled ‘ion’) presents ion energy density (red color bar) and laser intensity (green-blue color bar). The bottom panel (labeled ‘electron’) shows electron energy density (red color bar) and laser intensity (green-blue color bar). A lineout through the center of the simulation box displays the electric field component along the y-direction for the ion plots and the x-direction for the electron plots. Snapshots (a, b), (c, d), and (e, f) correspond to simulation times $t = 100\,\mathrm{fs}$, $130\,\mathrm{fs}$, and $140\,\mathrm{fs}$, respectively.}
		\label{fig:fig2}
	\end{figure}

	Figure~\ref{fig:fig2} captures the early interaction phase. At $t=100\,\mathrm{fs}$ [Fig.~\ref{fig:fig2}(a, b)], the laser pulse impinges on the target front. The ion energy density maps (panels a-1, b-1) reveal the initial interaction with the pre-plasma gradient. Unlike a sharp vacuum-solid interface, the target is initialized with a density gradient ($\delta = 0.1\,\mu\mathrm{m}$) rising to $n_e \approx 106\,n_{c}$. For the flat target [Fig.~\ref{fig:fig2}(a-2)], electron heating is visible within this gradient region. The lineout of $E_x$ (black/orange curve) indicates the onset of a sheath field, triggered by vacuum heating and resonance absorption mechanisms before the bulk plasma is significantly perturbed. Similarly, the annular sector target [Fig.~\ref{fig:fig2}(b-2)] exhibits distinct sheath formation along the concave inner surface, where the pre-plasma expansion into the void facilitates electron energization well before the laser peak reaches the geometric focus.
	
	By $t=130\,\mathrm{fs}$ [Fig.~\ref{fig:fig2}(c, d)], the interaction reaches peak intensity. In the flat target case, the highly overdense plasma acts as a relativistic mirror. The intense radiation pressure deforms the front surface, driving the critical density interface inward. The electron energy density map [Fig.~\ref{fig:fig2}(c-2)] shows a population of relativistic electrons traversing the $1\,\mu\mathrm{m}$ thick target and exiting the rear surface. This establishes the strong longitudinal sheath field ($E_x$) visible in the lineout, which is the primary driver for Target Normal Sheath Acceleration (TNSA).
	
	The annular sector target [Fig.~\ref{fig:fig2}(d)] displays distinct geometric effects. The laser intensity profile (green-blue map in d-1) shows the pulse entering the cavity and interacting with the expanding plasma lining the void. The curvature of the target focuses the reflected laser light, creating a distributed interaction volume. The electron density map [Fig.~\ref{fig:fig2}(d-2)] confirms that electrons are energized along the entire illuminated arc rather than at a single focal spot.
	
	By $t=140\,\mathrm{fs}$, the distinction between the two geometries becomes pronounced. For the flat foil, the laser pulse reflects and exits the simulation window [Fig.~\ref{fig:fig2}(e)]. Conversely, the annular geometry exhibits a strong \textit{cavity confinement effect}. Figure~\ref{fig:fig2}(f-1) reveals that a significant portion of the laser energy (indicated by the high-amplitude $E_y$ lineout) remains trapped within the target void ($x \approx 12$--$22\,\mu\mathrm{m}$). This trapped field undergoes multiple reflections off the conductive cavity walls, sustaining the electron heating process [Fig.~\ref{fig:fig2}(f-2)] significantly longer than the single-pass interaction observed in the flat geometry.
	
	\subsection{Late-Stage Evolution ($t=180$--$410\,\mathrm{fs}$)}
	
\begin{figure}
		\centering
		\includegraphics[width=0.9\linewidth]{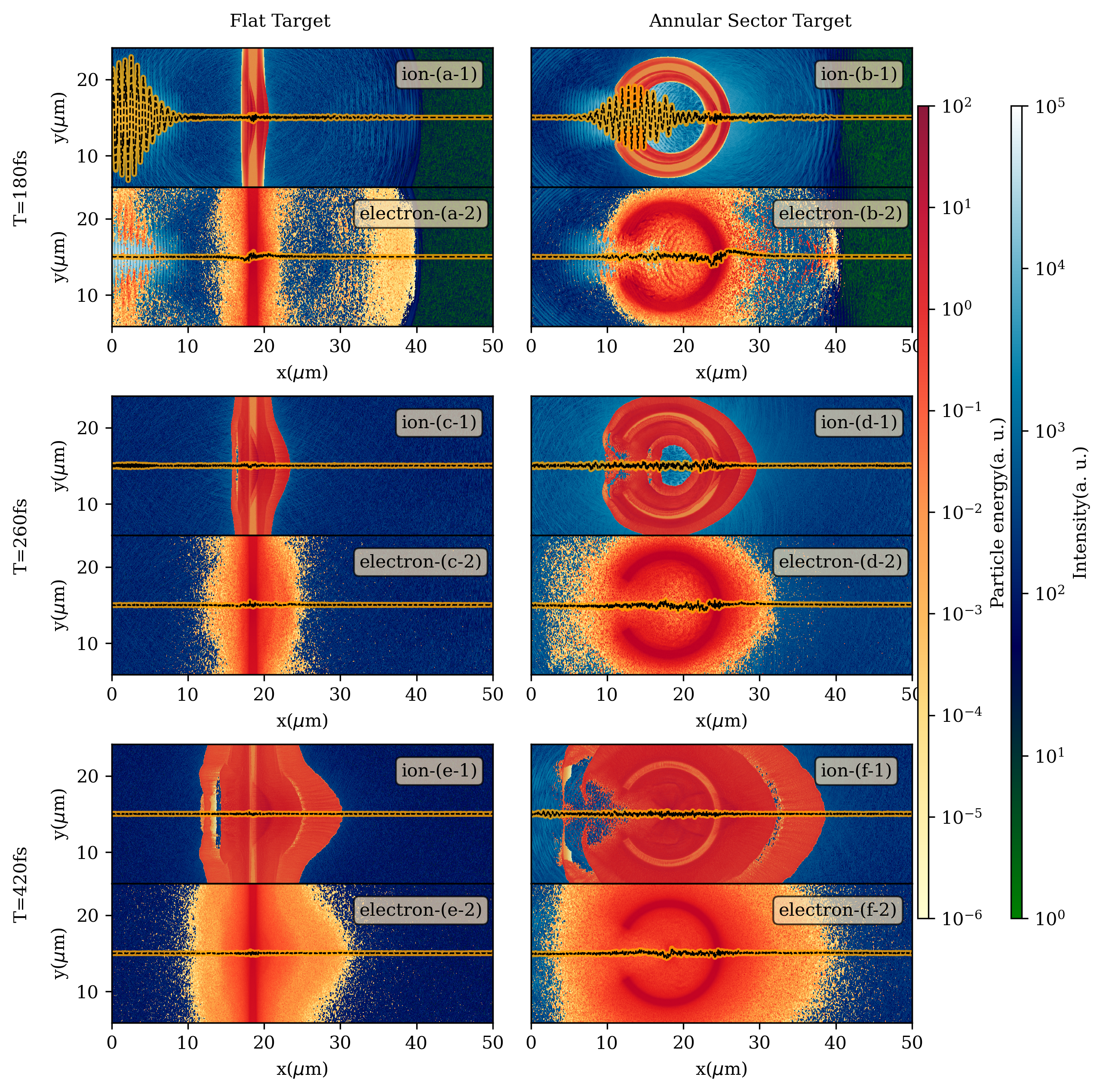}
		\caption{Comparison of ion and electron energy densities in laser-solid target interactions for flat and annular sector targets. The details are the same as in Fig. 2; snapshots (a, b), (c, d), and (e, f) correspond to simulation times $t = 180\,\mathrm{fs}$, $260\,\mathrm{fs}$, and $420\,\mathrm{fs}$, respectively.}
		\label{fig:fig3}
	\end{figure}

	Figure~\ref{fig:fig3} illustrates the relaxation and expansion phase ($t=180$--$410\,\mathrm{fs}$). This stage highlights the most significant difference between the two geometries: the transition from ballistic expansion to geometric convergence.
	
	For the flat target (left column), the interaction ceases abruptly once the laser passes. At $t=180\,\mathrm{fs}$ [Fig.~\ref{fig:fig3}(a)], the laser field intensity drops to zero. The system enters a ballistic expansion regime where the hot electron cloud spreads longitudinally into the vacuum, pulling ions with it. By $t=410\,\mathrm{fs}$ [Fig.~\ref{fig:fig3}(e)], the plasma has expanded significantly, but the electric field lineouts are flat, indicating that the accelerating forces have largely decayed and the electron population is undergoing adiabatic cooling.
	
	In contrast, the annular sector target maintains a dynamic electromagnetic environment long after the laser drive is off. At $t=180\,\mathrm{fs}$ [Fig.~\ref{fig:fig3}(b-1)], residual laser modes remain trapped within the cavity, bouncing between the reflective cavity walls. Crucially, the plasma lining the inner void—heated early by the pre-plasma interaction—begins to expand radially inward.
	
	This leads to a \textit{geometric plasma closure} effect. By $t=260\,\mathrm{fs}$ [Fig.~\ref{fig:fig3}(d)], the ion fronts from the opposing walls of the cavity are clearly visible converging towards the central axis ($y = 15\,\mu\mathrm{m}$), effectively filling the void. By $t=410\,\mathrm{fs}$ [Fig.~\ref{fig:fig3}(f)], the void is increasingly populated by dense, hot plasma. A key observation is the persistent electric field activity seen in the electron density lineout [Fig.~\ref{fig:fig3}(f-2)]. Unlike the flat target, the annular target exhibits fluctuating electrostatic fields in the center, indicating that the trapped electron population is still recirculating and interacting with the converging ion fronts. This sustained confinement prevents rapid cooling and maintains a localized energy density at the geometric focus ($x \approx 18\,\mu\mathrm{m}$).

	\begin{figure}
		\centering
		\includegraphics[width=0.9\linewidth]{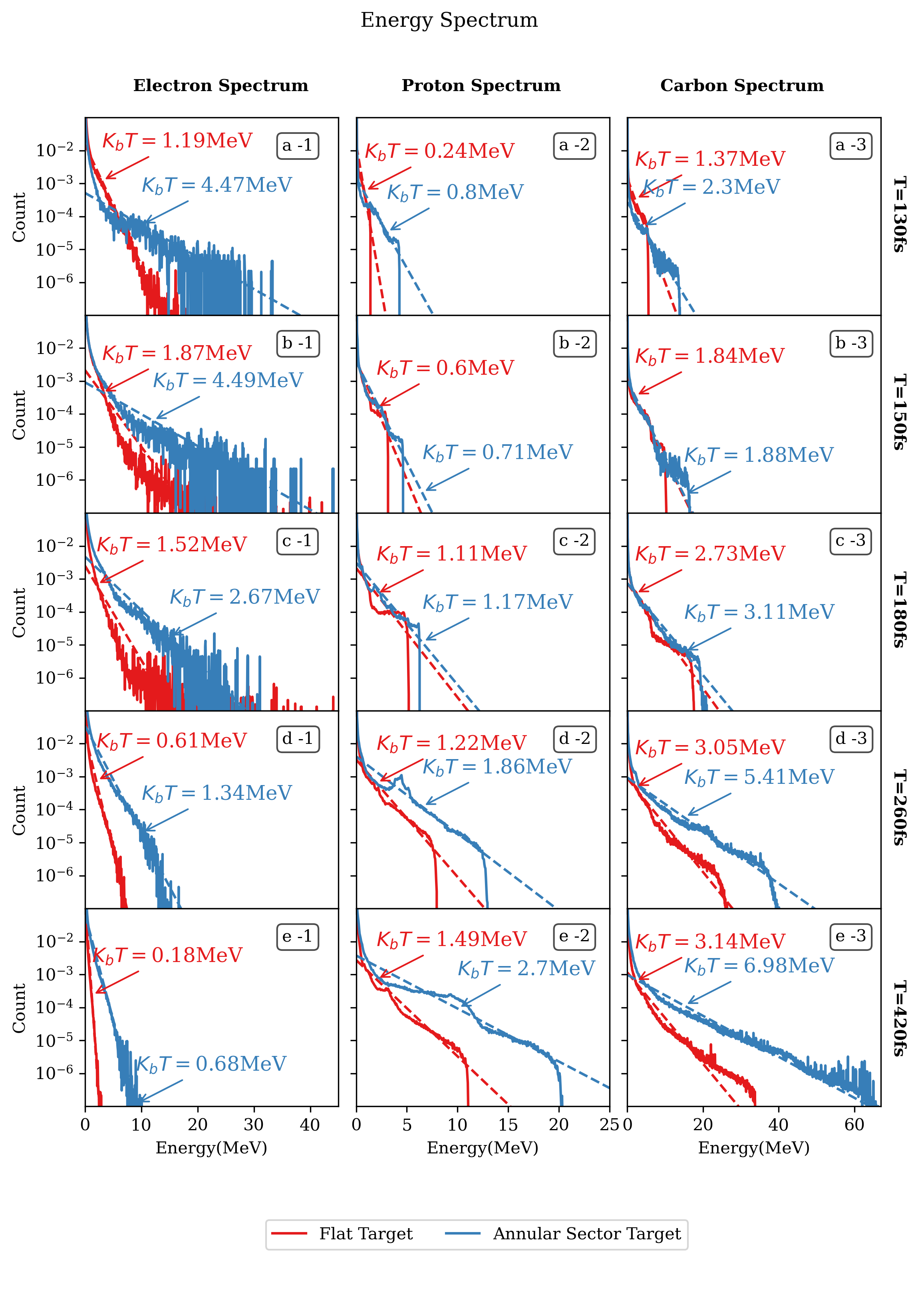}
		\caption{Temporal evolution of particle energy spectra. The columns correspond to electron (left), proton (middle), and Carbon ion (right) spectra for the flat (red curves) and annular sector (blue curves) targets. The rows represent simulation snapshots at $t = 130, 150, 180, 260, \text{and } 420\,\mathrm{fs}$. Solid lines depict the simulation data, while dashed lines indicate the Maxwellian fits used to estimate effective temperatures.}
		\label{fig:fig4}
	\end{figure}
	
	\section{Particle Energy Spectra and Acceleration Efficiency}
	
	To quantify the efficiency of the acceleration mechanisms described in the previous section, we analyzed the temporal evolution of the particle energy spectra. Figure~\ref{fig:fig4} presents the energy distributions for electrons, protons, and carbon ions at five distinct snapshots ranging from the peak of interaction ($t=130\,\mathrm{fs}$) to the late expansion phase ($t=420\,\mathrm{fs}$). Maxwellian distributions were fitted to the high-energy tails of these spectra to estimate the effective hot particle temperatures ($k_B T_{hot}$).
	
	\subsection{Electron Heating and Confinement}
	The electron energy spectra (Fig.~\ref{fig:fig4}, left column) provide direct evidence of enhanced laser absorption in the annular geometry. At the peak of interaction ($t=130\,\mathrm{fs}$), the flat target exhibits a hot electron temperature of $k_B T_e \approx 1.19\,\mathrm{MeV}$. In stark contrast, the annular sector target generates a significantly hotter electron population with $k_B T_e \approx 4.47\,\mathrm{MeV}$—approximately a fourfold increase. This enhancement confirms that the "cavity confinement" traps electrons within the oscillating laser field, allowing them to gain energy far beyond the limits of a single-pass interaction.
	
	Both target configurations exhibit a trend of decreasing electron temperature over time, though at different rates. For the flat target, the temperature drops rapidly to $0.18\,\mathrm{MeV}$ by $t=420\,\mathrm{fs}$. This decay is attributed to two main factors: (1) the rapid transfer of kinetic energy from hot electrons to the accelerating ions (establishing the sheath field), and (2) the loss of high-energy electrons that escape the simulation boundaries, a constraint inherent to fixed-window PIC simulations.
	
	However, the annular target mitigates this cooling. Even at $t=420\,\mathrm{fs}$, it maintains a temperature of $0.68\,\mathrm{MeV}$. The geometric closure of the cavity promotes electron recirculation, effectively keeping the hot electron population interacting with the ion front for a longer duration, thereby sustaining the acceleration field.
	
	\subsection{Ion Acceleration (Protons and Carbon)}
	The sustained high electron temperature in the annular geometry translates directly into enhanced ion acceleration via the TNSA mechanism.
	
	\textbf{Proton Acceleration:} The middle column of Fig.~\ref{fig:fig4} displays the proton spectra. At $t=130\,\mathrm{fs}$, the annular target already exhibits a higher cut-off energy ($\sim 5\,\mathrm{MeV}$) compared to the flat target. As time progresses, the persistent sheath field in the annular target continues to accelerate protons efficiently. By $t=420\,\mathrm{fs}$, the proton cut-off energy for the annular target reaches approximately $22\,\mathrm{MeV}$, whereas the flat target saturates around $12\,\mathrm{MeV}$. The effective proton temperature is similarly enhanced, reaching $2.7\,\mathrm{MeV}$ compared to $1.49\,\mathrm{MeV}$ for the flat case.
	
	\textbf{Carbon Acceleration:} The right column of Fig.~\ref{fig:fig4} confirms that this enhancement extends to heavier ion species. The Carbon spectra show a dramatic increase in maximum energy cut-off. At late times ($t=420\,\mathrm{fs}$), Carbon ions from the annular target reach energies exceeding $60\,\mathrm{MeV}$, compared to approximately $35\,\mathrm{MeV}$ for the flat target. The effective temperature of the accelerated Carbon bunch is exceptionally high in the annular case ($k_B T_C \approx 6.98\,\mathrm{MeV}$), indicating that the localized hot spot created by the geometric focusing of the plasma efficiently transfers energy to the heavier ions.
	
	In summary, the annular sector geometry yields a factor of $\sim 2$ increase in maximum ion energy for both protons and carbon compared to a standard flat foil. This gain is physically driven by the combined effects of optical field confinement (enhancing absorption) and geometric plasma focusing (sustaining the accelerating sheath fields).

	\begin{figure}
		\centering
		\includegraphics[width=0.9\linewidth]{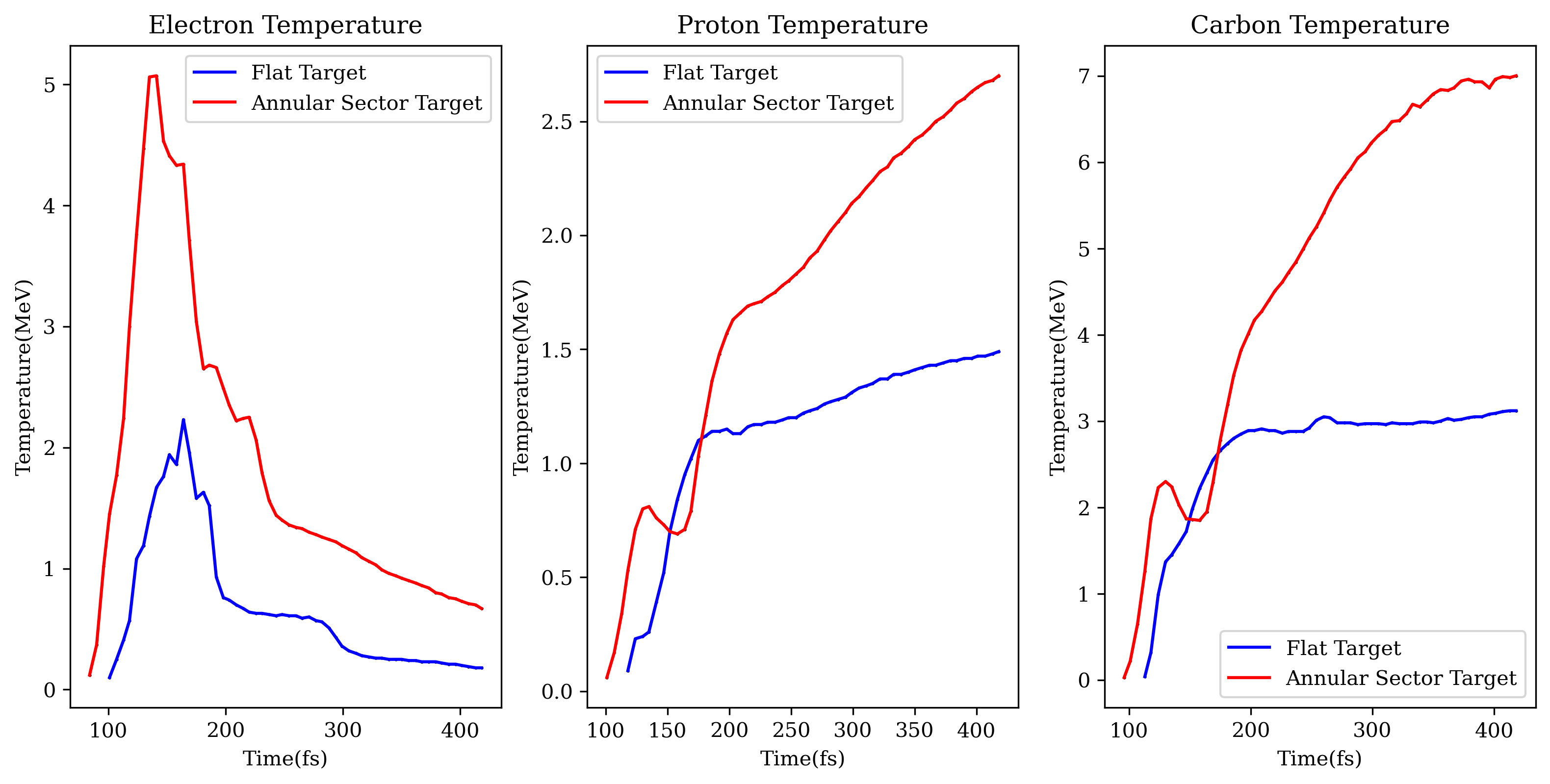}
		\caption{
			Temporal evolution of the effective particle temperatures for (a) electrons, (b) protons, and (c) Carbon ions. The blue curves correspond to the flat target, and the red curves correspond to the annular sector target. The annular geometry exhibits an earlier onset of heating and consistently higher peak temperatures. Notably, the Carbon temperature evolution (c) for the annular target reveals a step-like increase (at $t \approx 130\,\mathrm{fs}$ and $t \approx 160\,\mathrm{fs}$), correlating with the optical transit time across the cavity void.}
		\label{fig:fig5}
	\end{figure}

	\section{Temporal Dynamics of Particle Heating}
	
	To further elucidate the acceleration mechanisms, we tracked the temporal evolution of the effective temperatures for electrons, protons, and Carbon ions, as shown in Fig.~\ref{fig:fig5}.
	
	\subsection{Electron Heating Dynamics}
	The electron temperature evolution [Fig.~\ref{fig:fig5}(a)] confirms the "early onset" of interaction observed in the density maps. The heating for the annular sector target (red curve) begins at $t \approx 90\,\mathrm{fs}$, approximately $15\,\mathrm{fs}$ earlier than the flat target. This is attributed to the expanded pre-plasma within the target void, which facilitates earlier laser coupling.
	
	The peak electron temperature for the annular target reaches $T_e \approx 5.1\,\mathrm{MeV}$, significantly exceeding the flat target's peak of $2.2\,\mathrm{MeV}$. Following the peak, the electron temperature in the flat target decays rapidly due to adiabatic expansion. In contrast, the annular target maintains a higher residual temperature throughout the simulation. This supports the conclusion that cavity confinement prevents rapid cooling and sustains electron recirculation.
	
	\subsection{Multi-Stage Ion Acceleration}
	
	The ion temperature plots [Fig.~\ref{fig:fig5}(b, c)] reveal a fundamental difference in acceleration regimes. For the flat target (blue curves), both proton and Carbon temperatures rise initially but quickly saturate (e.g., Carbon saturates at $\sim 3\,\mathrm{MeV}$ by $t=200\,\mathrm{fs}$). This saturation occurs as the laser pulse departs and the sheath field decays. Conversely, the annular target (red curves) exhibits a continuous energy gain, reaching significantly higher final temperatures ($2.7\,\mathrm{MeV}$ for protons and $7.0\,\mathrm{MeV}$ for Carbon).
	
	Most importantly, the Carbon temperature evolution [Fig.~\ref{fig:fig5}(c)] provides a distinct signature of the \textit{cavity reflection mechanism}. The red curve exhibits a "step-like" heating profile characterized by three distinct phases. A primary heating phase occurs between $t=110$--$140\,\mathrm{fs}$, driving the temperature to $\sim 2\,\mathrm{MeV}$, followed by a distinct plateau lasting approximately $20\,\mathrm{fs}$ (from $t=140$ to $160\,\mathrm{fs}$). Subsequently, a secondary heating phase begins at $t \approx 160\,\mathrm{fs}$, driving the temperature toward $7\,\mathrm{MeV}$. This $20\,\mathrm{fs}$ delay corresponds precisely to the optical transit time across the $6\,\mu\mathrm{m}$ diameter of the cavity void ($t_{transit} \approx D/c \approx 20\,\mathrm{fs}$). This correlation strongly suggests that the reflected laser pulse (or the associated hot electron bunch) traverses the cavity and re-interacts with the opposing wall, delivering a "second kick" to the ion population. This multi-stage acceleration is the key factor responsible for the superior performance of the annular sector target.

	\begin{figure}
		\centering
		\includegraphics[width=0.9\linewidth]{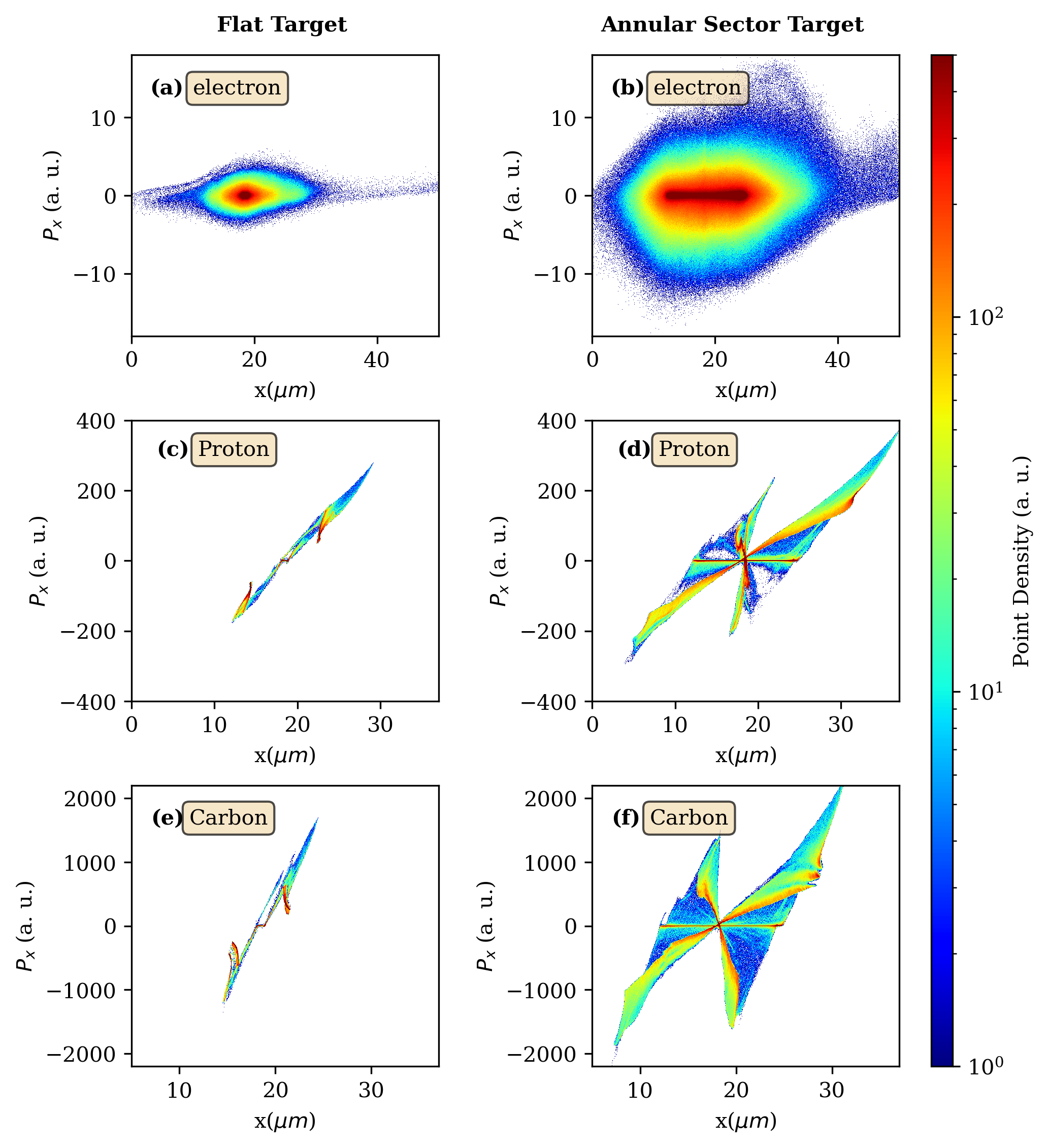}
		\caption{Longitudinal phase space ($P_x$ vs. $x$) for electrons (top), protons (middle), and Carbon ions (bottom) at the end of the simulation ($t=420\,\mathrm{fs}$). The left column displays the flat target, showing typical TNSA-driven expansion. The right column displays the annular sector target. Note the distinct "X-like" crossing structure in the ion phase spaces (d, f) at $x \approx 18\,\mu\mathrm{m}$, indicating the geometric focusing of particle beams converging from the upper and lower cavity walls.}
		\label{fig:fig6}
	\end{figure}

	\section{Phase Space Dynamics and Geometric Focusing}
	To visualize the momentum distribution of the accelerated bunches, we examined the longitudinal phase space ($P_x - x$) at the end of the interaction ($t=420\,\mathrm{fs}$), as shown in Fig.~\ref{fig:fig6}.
	
	For the flat target (left column), the electron phase space [Fig.~\ref{fig:fig6}(a)] shows a thermalized cloud centered at the target position, typical of thin-foil heating. The ion phase spaces [Fig.~\ref{fig:fig6}(c, e)] exhibit a classic TNSA linear expansion structure, where momentum increases linearly with distance from the target rear surface. This indicates a divergent beam expanding into the vacuum.
	
	In contrast, the annular sector target (right column) reveals a highly structured acceleration dynamic driven by the cavity geometry. The electron cloud [Fig.~\ref{fig:fig6}(b)] is significantly hotter and spatially broader, filling the entire void region ($x=10$--$25\,\mu\mathrm{m}$), with a distinct "hot core" at $x \approx 18\,\mu\mathrm{m}$ that confirms the accumulation of energetic electrons at the geometric focus. regarding the ions, the proton and Carbon phase spaces [Fig.~\ref{fig:fig6}(d, f)] display a striking "X-like" crossing pattern centered at $x \approx 18\,\mu\mathrm{m}$. This structure arises from the convergence of two distinct ion populations accelerated from the upper and lower arms of the "C-shaped" cavity. As these bunches converge towards the axis, they cross in phase space, creating a localized region of extremely high momentum density. This focusing effect explains the enhanced cut-off energies observed in the spectra (Fig.~\ref{fig:fig4}), as the converging sheath fields superimpose at the focal point.
	
	\begin{figure}
		\centering
		\includegraphics[width=0.7\linewidth]{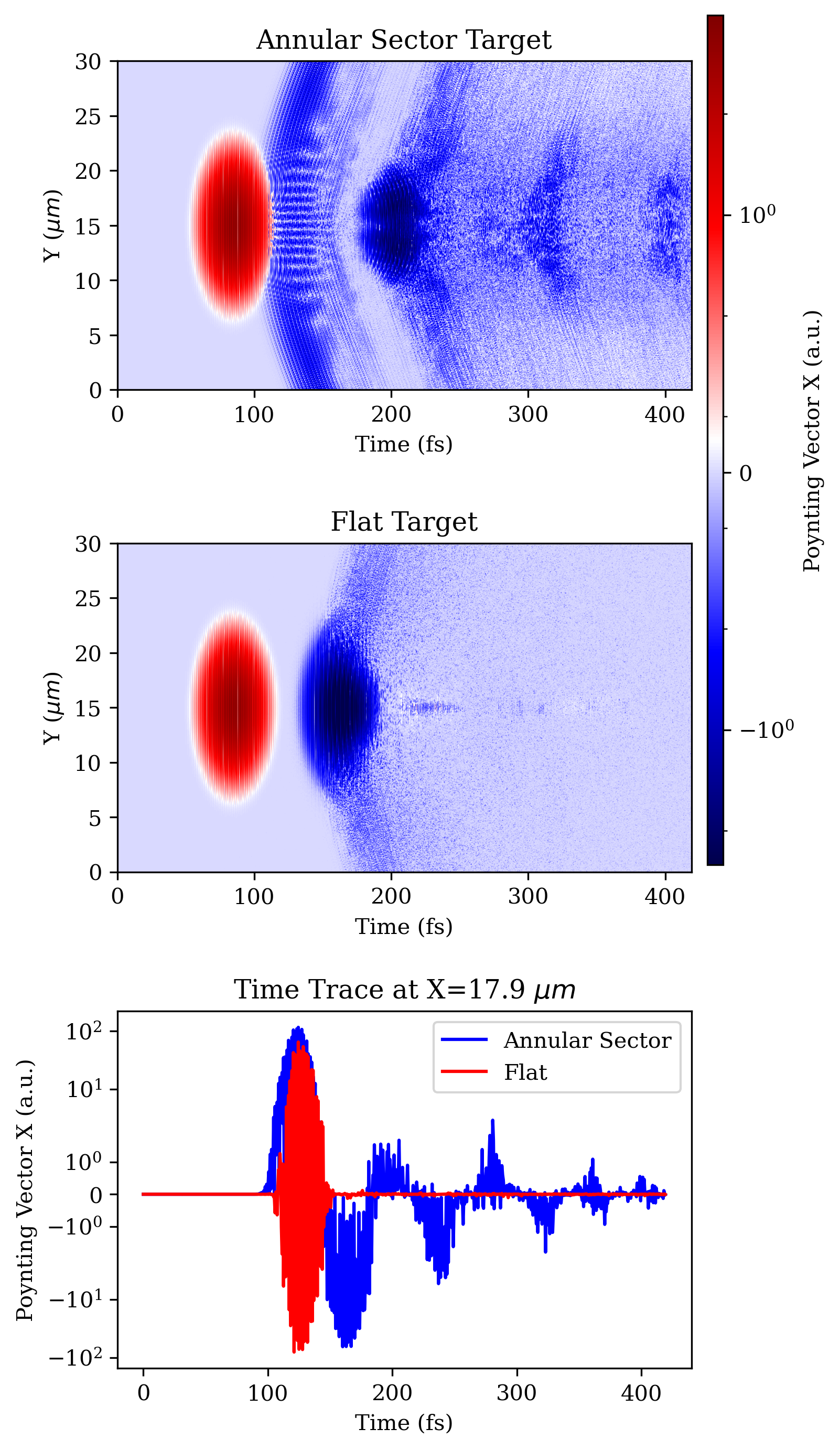}
		\caption{Spatiotemporal evolution of the longitudinal Poynting vector ($S_x$). (Top/Middle) Streak images showing $S_x$ along a transverse line probe at the cavity entrance ($x=5\,\mu\mathrm{m}$) over time. The flat target (middle) shows a single reflection event. The annular target (top) shows multiple distinct reflection bands. (Bottom) Time trace of the Poynting vector at a point probe near the target center ($x=17.9\,\mu\mathrm{m}$). The flat target signal (red) vanishes after $t=150\,\mathrm{fs}$. The annular target signal (blue) persists with high-amplitude oscillations for $>300\,\mathrm{fs}$, confirming long-term optical confinement.}
		\label{fig:fig7}
	\end{figure}

	\section{Electromagnetic Field Confinement}
	
	The sustained heating and multi-stage acceleration observed in the annular target rely on the premise that the laser pulse is optically trapped. To verify this, we monitored the longitudinal Poynting flux ($S_x$). Figure~\ref{fig:fig7} (top panels) displays streak images taken at the target entrance. For the flat target (middle panel), the interaction is ballistic: the incident pulse arrives at $t \approx 80\,\mathrm{fs}$, reflects off the critical surface, and travels back through the probe plane as a single return signal (blue region at $t \approx 150\,\mathrm{fs}$), leaving the region quiet thereafter. In contrast, the annular target (top panel) exhibits a complex, multi-banded reflection pattern. Following the primary reflection, distinct secondary and tertiary signals appear at late times ($t > 200\,\mathrm{fs}$), indicating that light is leaking out of the cavity only after undergoing multiple internal reflections.
	
	This confinement is most clearly quantified by the local time trace at the target center ($x=17.9\,\mu\mathrm{m}$), shown in the bottom panel of Fig.~\ref{fig:fig7}. While the flat target signal (red curve) shows the laser passing through the focal plane and vanishing completely by $t=150\,\mathrm{fs}$, the annular target signal (blue curve) reveals a starkly different behavior. Although it initially matches the flat target profile, it continues to oscillate with significant amplitude long after the main pulse has passed, exhibiting distinct high-intensity bursts at $t \approx 200\,\mathrm{fs}$, $280\,\mathrm{fs}$, and $350\,\mathrm{fs}$.
	
	These periodic bursts correspond to the laser pulse bouncing between the cavity walls and returning to the center. To quantify the cumulative effect of this confinement, we calculated the total laser energy absorption by integrating the Poynting vector flux over the 2D simulation domain. This analysis reveals a substantial enhancement in coupling efficiency: whereas the flat target absorbs approximately $16\%$ of the incident laser energy, the annular sector target achieves an absorption of $49\%$. This three-fold increase in energy uptake, driven by the persistent "optical cavity" mechanism, directly explains the continuous rise in ion temperatures reported in Fig.~\ref{fig:fig5}.

	\section{Conclusion}
	
	We have performed a comprehensive 2D PIC simulation study comparing flat and annular sector (C-shaped) target geometries for TNSA-driven ion acceleration. Our findings unequivocally reveal a substantial advantage of the annular sector target, driven by a synergy between electromagnetic confinement and geometric plasma focusing.
	
	The key mechanism for this enhancement lies in the target's ability to act as an optical cavity. Analysis of the Poynting vector evolution [Fig.~\ref{fig:fig7}] confirmed that the annular geometry effectively traps the laser pulse, causing it to undergo multiple reflections within the void. This confinement extends the interaction lifetime significantly, maintaining strong oscillating fields for over $300\,\mathrm{fs}$—long after the pulse would have dissipated in a flat target scenario.
	
	This sustained interaction leads to superior energy coupling. The annular target achieved a peak electron temperature of $T_e \approx 5.1\,\mathrm{MeV}$, more than double the peak observed in the flat target ($2.2\,\mathrm{MeV}$). Furthermore, the temporal evolution of Carbon ion temperatures [Fig.~\ref{fig:fig5}] revealed a "step-like" heating signature, providing direct evidence of multi-stage acceleration driven by the transit of the reflected laser pulse across the cavity.
	
	These dynamics translate into a potent enhancement of ion beam quality. The longitudinal phase space analysis [Fig.~\ref{fig:fig6}] uncovered a distinct "X-like" focusing structure for the annular target, where ion bunches accelerated from the converging cavity walls superimpose at the geometric center. Consequently, the maximum proton cut-off energy reached $\approx 22\,\mathrm{MeV}$ (compared to $12\,\mathrm{MeV}$ for the flat target), while Carbon ions exceeded $60\,\mathrm{MeV}$—nearly a two-fold increase in maximum energy.
	
	In summary, our results establish that modifying the target surface curvature to form a semi-enclosed cavity is a critical parameter for optimizing TNSA efficiency. By coupling the "optical trap" effect with geometric plasma focusing, the annular sector target offers a promising pathway for designing compact, high-energy ion sources. While fabricating such micro-structured targets presents experimental challenges, recent advances in high-precision micro- and nano-machining suggest that realizing these geometries is increasingly feasible. Future work will focus on optimizing the cavity dimensions and opening angle to maximize beam collimation and yield for practical applications.

		\section*{Funding}
		This work is based upon research funded by Iran National Science Foundation (INSF) under project No.4036338.
		\section*{Acknowledgments}
		The authors would like to thank Dr. Babak Zare Rameshti for providing access to high-performance computing (HPC) facilities, which were essential for the simulations performed in this study. 
	\section*{Data availability}
		The data that support the findings of this study are available from the corresponding author upon reasonable request.

	\bibliographystyle{unsrt}

	\bibliography{sampleCopy}

\end{document}